\documentclass[11pt,letterpaper]{article}
\usepackage[T1]{fontenc}
\usepackage[utf8]{inputenc}
\usepackage{mathpazo}
\usepackage{microtype,graphicx,amsmath,amssymb,booktabs,tabularx,array}
\usepackage[letterpaper,left=0.82in,right=0.82in,top=0.83in,bottom=0.84in]{geometry}
\usepackage{enumitem,caption,multicol,xurl,needspace}
\usepackage[section]{placeins}
\usepackage{xcolor}
\definecolor{linkblue}{HTML}{214C84}
\usepackage[colorlinks=true,urlcolor=linkblue,linkcolor=linkblue,bookmarksnumbered=true]{hyperref}
\hypersetup{pdftitle={The Rebundling of Freelance Work in the AI Age},pdfauthor={Caroline Chin, Shagun Kala, Nadav Kunievsky, Teng Liu, Alex Martin Richmond, Carl Shan}}
\counterwithin{figure}{section}
\counterwithin{table}{section}

\newcommand{\exref}[1]{\hyperref[fig:#1]{Exhibit~\ref*{fig:#1}}}
\newcommand{\methodex}[2][]{\methodhead{\exref{#2}. \nameref*{fig:#2}#1}}
\setlist[itemize]{nosep,leftmargin=1.05em}

\newcommand{\methodhead}[1]{\par\Needspace{4\baselineskip}\medskip\noindent\textbf{#1}\par\nopagebreak}
\newenvironment{classifierprompt}{%
  \par\linespread{1}\ttfamily\fontsize{10.5}{13}\selectfont
  \microtypesetup{expansion=false}\raggedright
  \setlength{\parindent}{0pt}\setlength{\parskip}{4pt}%
  \hyphenpenalty=10000\relax
}{\par}
\begin{document}
\pdfbookmark[0]{The Rebundling of Freelance Work in the AI Age}{title}
\begin{center}
{\fontsize{23}{27}\selectfont\scshape The Rebundling of Freelance Work\\in the AI Age\par}
\vspace{1.4em}
{\large Caroline Chin\textsuperscript{1}\qquad Shagun Kala\textsuperscript{2}\qquad Nadav Kunievsky\textsuperscript{2}\\[3pt]
Teng Liu\textsuperscript{2}\qquad Alex Martin Richmond\textsuperscript{1}\qquad Carl Shan\textsuperscript{1}\par}
\vspace{.8em}
\textsuperscript{1}OpenAI\qquad\textsuperscript{2}Upwork\par
\vspace{.3em}
November 2026
\end{center}
\vspace{1.1em}
\begin{center}\begin{minipage}{0.87\linewidth}
\begin{center}\textbf{Abstract}\end{center}

This report studies how freelance work is changing in the AI era. We compare the activities people
undertake in ChatGPT with the work clients request on Upwork and examine how clients combine tasks
into projects. AI-related postings request slightly more tasks on average than other postings. Even
among projects with similar numbers of tasks, AI-related postings bring together tasks that clients
less often requested together before the generative AI era.

\end{minipage}\end{center}

\vspace{.6em}

\begin{center}\begin{minipage}{.87\linewidth}\small

\textbf{Key findings}\par\medskip

\noindent\textbf{AI use and freelance demand have different activity mixes.} ChatGPT activities skew toward writing and advice, while Upwork postings lean toward technical implementation and visual production. This pattern is consistent with businesses continuing to hire freelancers for work that requires implementation, production, and delivery.\par\smallskip

\noindent\textbf{The relationship between direct AI use and changes in freelance demand is not uniform: some activities gaining share on Upwork are common in ChatGPT, while others are not.} There is not a simple substitution pattern between Upwork tasks and those common on ChatGPT.  This contrast suggests a continuing role for workers who provide technical implementation, creative production, and finished deliverables.\par\smallskip

\noindent\textbf{AI-related work is becoming more diverse.} Work within AI-related postings is becoming less concentrated in a few activities. Among postings with at least two tasks, the average share of historically uncommon task pairs is higher for AI-related work.  AI is changing both the breadth and mix of work that clients seek to hire.\par\smallskip

\noindent\textbf{AI-related postings carry larger project budgets.} Posted fixed-price budgets are approximately 52 percent higher in AI-related postings than for non-AI work after accounting for work subcategory and posting month. The value of AI-related freelance work may lie partly in delivering broader projects that combine many capabilities.\par\smallskip

\end{minipage}\end{center}

\vfill\noindent{\footnotesize Authorship listed in alphabetical order.}

\clearpage

\section{Introduction}

\subsection{How AI is changing work}

Public debate about AI and work often centers on a single question: how many jobs will AI replace? Yet,
technological change can reshape tasks and workflows well before occupations disappear or new jobs enter
official statistics. Understanding AI's labor-market effects holistically requires examining how the
content and organization of work are changing today.

The first report in \href{https://cdn.openai.com/pdf/work-at-the-frontier-report.pdf}{OpenAI Economic Research's ``Work at the Frontier'' series} provides evidence of one way users are starting to broaden their work activity. It finds that 43.5 percent of work-related ChatGPT messages assigned to an occupation-specific activity involve tasks traditionally associated with an occupation other than the user's own. This ``task crossover'' suggests that AI is expanding the range of activities people attempt beyond established occupational boundaries. But ChatGPT data alone do not show how demand for freelance work is changing, or how clients are combining different activities to form projects.

This is important to understand, as technical capability does not on its own determine labor-market
outcomes. When AI can perform part of a job, employers and firms may respond in different ways. Some
activities may be performed directly with AI rather than purchased from another person. In other cases,
lower costs may expand demand, or complementary work may become more valuable---for example, integrating
AI into systems, evaluating AI outputs, and managing complex projects.

Upwork Marketplace is a global platform that connects businesses with on-demand access to highly skilled
independent talent. The platform data provide a demand-side view of this adjustment. Job postings show
which activities clients seek, how those activities are bundled into projects, and the budgets or rates
clients attach to them. Freelance markets may reveal these changes relatively quickly because clients can
readily post emerging work, purchase one component of a larger role, or seek combinations of tasks that
previously belonged to different specialists.

This is represented in prior Upwork research, which suggests that AI is \href{https://www.upwork.com/research/redesigning-work-through-ai}{reorganizing demand} toward new forms of work, with discrete \href{https://www.upwork.com/research/research-future-workforce-index-2026}{AI-execution tasks} becoming more common but less lucrative, while demand is growing for more complex AI-augmented services, including work that involves supervising, reviewing, and correcting AI-generated outputs.

Upwork provides a useful setting for comparison with ChatGPT because its projects involve work that is
already remote and digitally mediated. This makes it possible to compare activities people undertake
directly with AI with digitally delivered work that clients seek from freelancers. The datasets are not
directly matched, and neither represents the labor market as a whole. We therefore interpret their
similarities and differences as descriptive evidence about the organization of work, with limits on what
they can establish about substitution between direct AI use and paid freelance services.

This report combines evidence from ChatGPT use and Upwork job postings to examine two dimensions of
change. First, we compare the activities people undertake directly with ChatGPT with the work clients
purchase from freelancers, and examine how the mix and stated value of Upwork demand are changing.
Second, we study how activities are organized within job postings---whether AI-related projects contain
broader task bundles and whether they combine tasks in ways that had less precedent before the
generative-AI era.

We find both similarities and differences in the activities people undertake with ChatGPT and those
clients seek from freelancers. Within AI-related Upwork Marketplace postings, demand is spreading across
a broader range of activities and shifting toward integration (e.g., integrating AI APIs into
applications), multimodal production (e.g., creating AI-generated video and producing motion graphics),
and applied business workflows (e.g., marketing automation and social-media campaign management).
AI-related jobs contain slightly more activities on average, but they combine those activities in novel
ways---even when compared with non-AI jobs of similar breadth in the same work subcategory and month.
AI-related postings also carry higher stated project budgets and hourly rates, even within narrowly
defined work categories. Together, the results provide marketplace evidence that AI is changing not only
which activities people undertake, but also how clients assemble work into projects.

\subsection{Related research and contribution}

A longstanding literature argues that the labor-market effects of technological change are best
understood by examining tasks rather than occupations. One common idea in this literature is that
technology can substitute for workers in some activities while increasing the value of complementary
human inputs, and the creation of new tasks can offset some of the resulting displacement
(\hyperlink{ref2}{Autor, 2015}; \hyperlink{ref1}{Acemoglu and Restrepo, 2019}; \hyperlink{ref3}{Autor et
al., 2024}). This perspective suggests that the effects of generative AI may first appear as changes in
what workers do and how activities are organized.

Research on generative AI documents both broad technical exposure and uneven effects in practice.
\hyperlink{ref10}{Eloundou et al. (2024)} estimate that large language models could affect tasks across a
wide range of occupations. Experimental and workplace studies find that generative AI can improve
productivity and output quality in professional writing and customer support, with especially large gains
for less experienced customer-support workers (\hyperlink{ref15}{Noy and Zhang, 2023};
\hyperlink{ref5}{Brynjolfsson, Li, and Raymond, 2025}). At the same time, the benefits depend on the
task: \hyperlink{ref7}{Dell'Acqua et al. (2026)} find that AI improves performance on tasks within its
capabilities but can reduce accuracy on tasks outside this ``jagged technological frontier.''

Recent theory emphasizes that jobs are endogenous bundles of tasks rather than fixed occupational
containers, with boundaries set by the tradeoff between specialization and coordination costs
(\hyperlink{ref18}{Rosen, 1978}; \hyperlink{ref4}{Becker and Murphy, 1992}; \hyperlink{ref16}{Ocampo,
2022}). Because AI can change the costs of performing tasks, coordinating across them, and transferring
context between them, it can split and rebundle jobs even when no task is fully automated
(\hyperlink{ref8}{Demirer et al., 2026}; \hyperlink{ref11}{Gans, 2026}), with effects that depend on how
easily tasks can be separated from the jobs they sit in (\hyperlink{ref12}{Garicano, Li, and Wu, 2026}).
To the best of our knowledge, direct empirical evidence on how AI is reshaping task bundles within real
projects remains limited.

Evidence from online labor markets points to both demand-side disruptions and worker adaptation.
\hyperlink{ref13}{Hui, Reshef, and Zhou (2024)} find declines in employment and earnings among
freelancers in categories more exposed to generative AI. \hyperlink{ref20}{Yiu et al. (2026)} show that
more-exposed freelancers also respond by shifting away from their previous domains and directing their
bids toward higher-value contracts, although domain-specific human capital can constrain this
repositioning. This report complements that work by focusing on the organization of the work being
purchased, using job-posting data from an online labor marketplace.

Additionally, technological change can reorganize work not only by creating or eliminating individual
tasks, but also by changing the boundaries that determine which tasks people do together. Which tasks are
bundled into a job reflects, in part, a classic tradeoff in the organization of work.
\hyperlink{ref18}{Rosen (1978)} emphasizes how substitution possibilities across different kinds of labor
shape the division of labor, while \hyperlink{ref4}{Becker and Murphy (1992)} show that greater
specialization must be balanced against the costs of coordinating complementary activities across
workers. Job boundaries therefore reflect both the gains from dividing work more finely and the costs of
separating activities that must ultimately be combined in production. AI can shift this tradeoff by
changing the relative costs of performing particular tasks, coordinating across them, and transferring
context from one task or worker to another.

Recent research further examines how AI can change the way tasks are grouped into jobs.
\hyperlink{ref8}{Demirer et al. (2026)} model firms as choosing how production steps are grouped into
tasks and jobs, as AI changes which sequences of work can be performed together. \hyperlink{ref11}{Gans
(2026)} shows that AI can induce task rebundling even without fully automating an activity by changing
the costs of transferring context across task boundaries. \hyperlink{ref12}{Garicano, Li, and Wu (2026)}
similarly show that the effects of AI depend on how easily tasks can be separated from the jobs in which
they are embedded. When tasks form a weak bundle, AI can pull one component out and narrow the remaining
human job, whereas stronger task complementarities can keep the existing bundle intact.

This idea also builds on a broader task-based theory of occupational boundaries. In
\hyperlink{ref16}{Ocampo (2022)}, workers have discrete skill types while tasks vary continuously in
their skill requirements, and occupations emerge endogenously as bundles of tasks assigned to workers.
Technological change alters comparative advantage and task prices, which can change the optimal
assignment of tasks across workers. Existing bundles can therefore break apart and tasks can be
reassigned and regrouped, redrawing the boundaries of occupations. Another aim of this report is to
understand how work is rebundling and reorganizing in response to AI.

\section{AI use and freelance hiring point to a changing composition of work}

\subsection{Upwork and ChatGPT comparison}

ChatGPT conversations and Upwork job posts offer complementary views of how work gets done. ChatGPT data
show which activities people undertake directly with a general-purpose AI system, while Upwork data show
which activities clients seek freelancers to perform. Comparing their distribution across broad work
areas reveals where direct AI use and paid freelance demand align and where they differ.

\exref{activity-shares} shows that how people use ChatGPT and the work that employers solicit on Upwork exhibit markedly different activities. This contrast suggests a division between activities people frequently undertake through a general-purpose AI interface and those for which clients seek specialized remote workers. Direct ChatGPT use is more concentrated in text-based and advisory activities, while Upwork postings are more concentrated in technical implementation, visual production, and work that often requires specialized tools, domain expertise, or delivery of a finished product.

Content writing and professional and business writing account for 14.5 percent and 10.2 percent of
sampled ChatGPT conversations, respectively, compared with approximately 1.3 percent and 0.4 percent of
Upwork job posts. Customer service, financial planning, and market research are also relatively more
prominent in ChatGPT use. By contrast, graphic and presentation design, video and animation,
quality-assurance testing, mobile development, digital marketing, and web development represent
substantially larger shares of Upwork postings than of ChatGPT conversations.

\begin{figure}[p]
\centering
\caption{Activity shares in ChatGPT and Upwork}\label{fig:activity-shares}
\includegraphics[trim=0 0 0 125bp,clip,width=\linewidth,height=0.79\textheight,keepaspectratio]{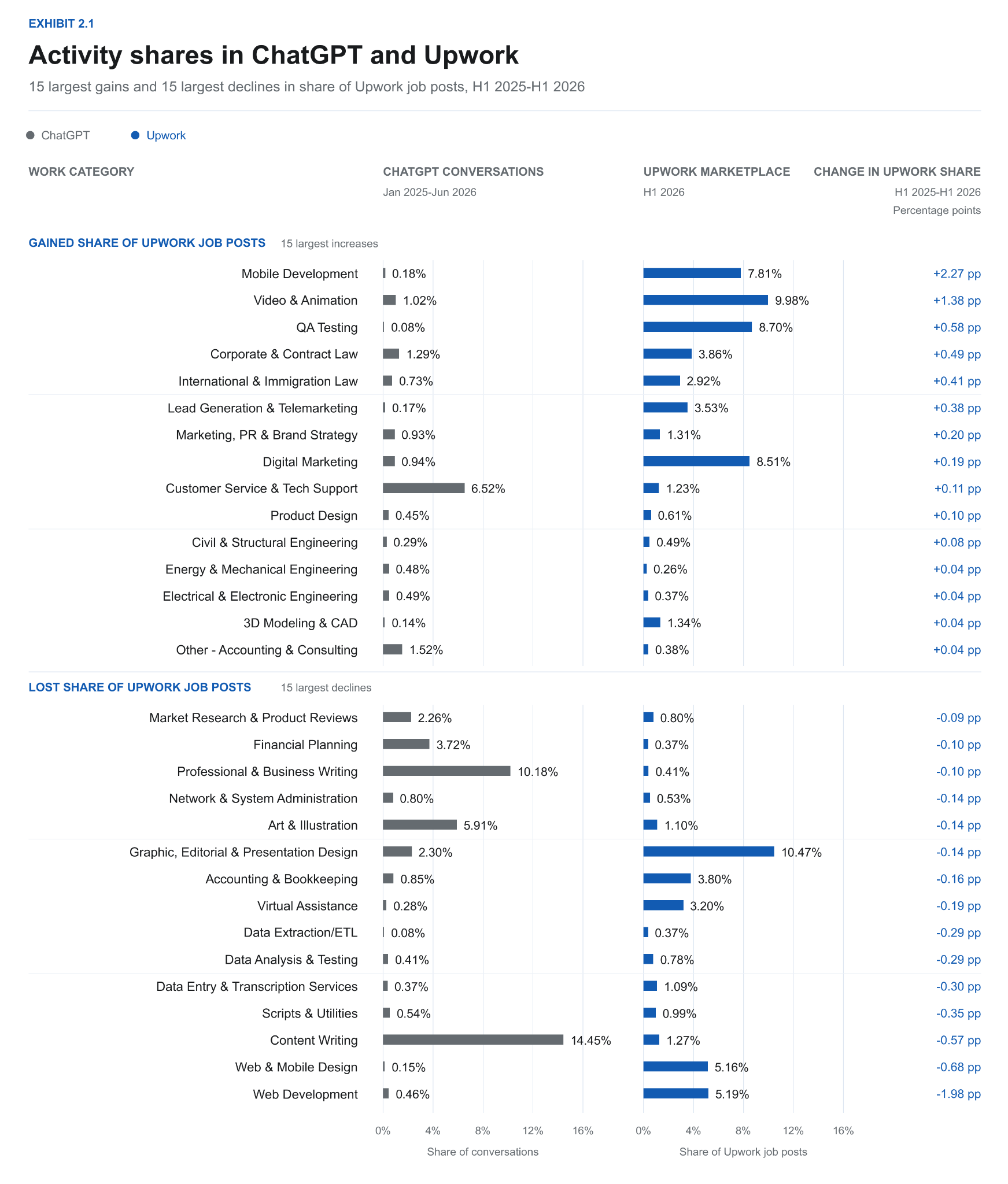}
\par\smallskip
\begin{minipage}{\linewidth}\footnotesize\raggedright
\textit{Note.} ChatGPT bars show the percentage of over 36,000 sampled personal or work-related conversations assigned to each work subcategory in the Upwork taxonomy using a privacy-preserving classifier. Schoolwork, unclear-purpose conversations, and conversations without an Upwork subcategory match are excluded. Upwork bars show each category's percentage of all Marketplace job posts in H1 2026. Categories not shown are included in the totals used to calculate both sets of percentages. Change in share is the H1 2026 Upwork share minus the H1 2025 share, in percentage points. The figure shows the 15 largest gains and 15 largest declines in share of Upwork job posts, H1 2025--H1 2026.

\end{minipage}
\end{figure}

Next, we examine whether the relative intensity of AI use across activities is related to changes in the
composition of demand for freelance work on Upwork Marketplace. We treat Upwork Marketplace job postings
in H1 2023 as a baseline for how demand for remote work was distributed across activities. ChatGPT
provides a way to use AI to perform many activities, but some may still prefer to hire human workers for
many reasons, including if AI capabilities are lacking on their specific tasks. We therefore define AI
intensity as the ratio of an activity's share of ChatGPT use in H2 2025 to its share of Upwork postings
in H1 2023. The measure captures the extent to which activity is tilted toward AI relative to remote
labor. Higher AI intensity identifies activities that are disproportionately prevalent in ChatGPT use
relative to their initial importance in the remote-work market. Lower AI intensity identifies activities
that remain relatively more concentrated in work delegated to remote workers.

The measure also captures revealed preferences in where people choose to use ChatGPT relative to the
initial mix of activities clients sought to delegate to freelancers. Higher intensity may indicate
activities where ChatGPT offers a useful and accessible way to perform the work. Lower intensity may
reflect greater reliance on specialized expertise, tools, or delivery of a finished product. These
choices reflect how AI's capabilities interact with users' needs and the practical costs of completing
work.

\exref{chatgpt-intensity} shows how each of 63 matched Upwork work subcategories changed as a share
of marketplace postings from H1 2023 to H2 2025. Subcategories with higher ChatGPT intensity tended
to lose marketplace share, while those with lower intensity more often gained share.

Content writing was particularly prominent in ChatGPT use, while freelance postings for it declined
sharply. It accounted for roughly three times the share of ChatGPT activity in H1 2025 as of Upwork
postings in H1 2023. Over that two-year period, content-writing postings fell at more than twice the rate
of Upwork postings overall. For activities such as content writing, people may use ChatGPT directly to
complete drafts of blog posts, marketing copy, and product descriptions that they would otherwise
commission from a freelancer. Video \& Animation illustrates the other side of this pattern, with lower
AI intensity and postings that grew by about a third relative to the overall marketplace. Here, clients
may continue to seek freelancers for video editing and animation production. Together, these examples
suggest how direct AI use may be shifting the boundary between work people perform themselves and work
they purchase from freelancers.

Of the 63 subcategories, 43 (68.3 percent) either had lower ChatGPT intensity and gained Upwork
marketplace share or had higher intensity and lost share. This suggests a shift in the allocation of work:
activities people readily undertake through ChatGPT may face weaker demand for freelance services, while
activities requiring specialized implementation or production---frontier tasks, many of which are not
currently as common on ChatGPT---may retain a larger role for freelancers.

The other 20 subcategories show why a simple substitution story explains only part of the pattern. Six categories (9.5 percent) combined higher ChatGPT intensity with gains in Upwork marketplace share;
the remaining 14 categories (22.2 percent) combined lower ChatGPT intensity with losses in Upwork marketplace
share. The presence of activities in these bins suggests that AI may complement freelance services in
some activities, while changes in client needs, production processes, and specialized software also shape
which activities gain or lose marketplace share.

\begin{figure}[htbp]
\centering
\caption{ChatGPT intensity and relative change in Upwork category share}\label{fig:chatgpt-intensity}
\includegraphics[width=\linewidth,height=0.79\textheight,keepaspectratio]{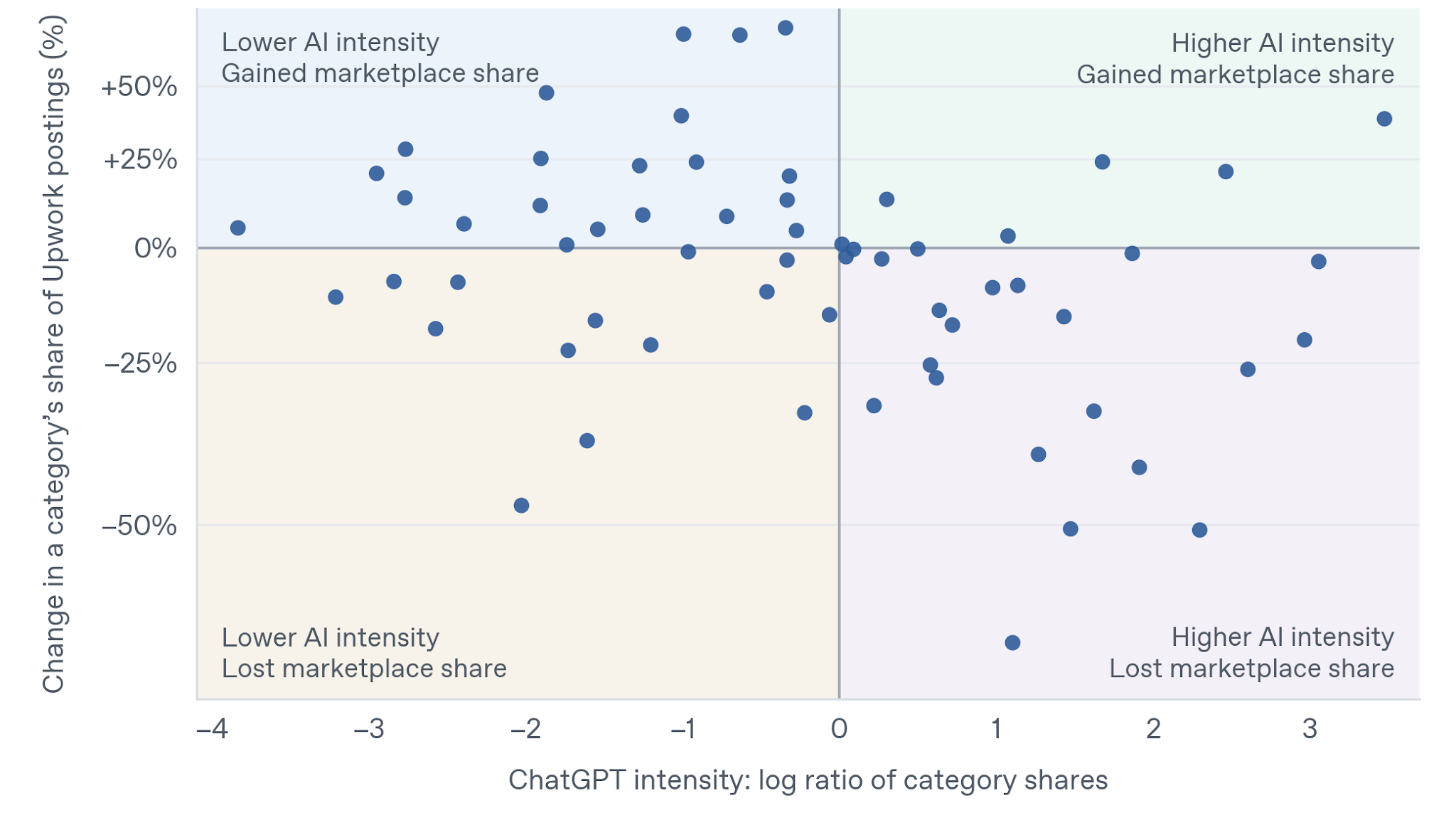}
\par\smallskip
\begin{minipage}{\linewidth}\footnotesize\raggedright
\textit{Note.} Each point represents one of 63 matched work subcategories in the Upwork taxonomy. The horizontal axis is the natural log of the category's share of ChatGPT use in H2 2025 divided by its share of Upwork job posts in H1 2023. The vertical axis shows the percentage change in the category's share of Upwork postings from H1 2023 to H2 2025, on a logarithmic scale; 0\% indicates unchanged marketplace share. The vertical divider at AI intensity 0 marks equal category shares in ChatGPT and baseline Upwork postings.

\end{minipage}
\end{figure}

We next look inside individual Upwork postings to identify the tasks clients request. For this
analysis, Upwork groups more than 3,000 detailed activity descriptions into more than 400
standardized tasks. These activity descriptions were
generated using an LLM to identify the specific activities involved in each service offered on Upwork,
drawing on job descriptions and associated skills. We group similar activity descriptions using semantic
similarity, LLM-assisted comparisons, and manual review. We refer to these standardized groups of closely
related activity descriptions as tasks. A job's task bundle consists of the distinct tasks it contains.
We use the same task definitions across all periods to examine how the composition of AI-related work
changes over time (see \hyperlink{task-construction}{Task construction in Appendix~\ref*{app:methodology}}).

\exref{activity-shares} shows how demand shifted across broad categories of freelance work between H1 2025 and H1 2026. \exref{task-composition} looks inside AI-related projects, tracing changes in their task composition between 2023 and 2025. This more detailed view reveals which activities gained importance as clients began incorporating AI into their work. The 2023--2025 task comparison captures an initial phase of generative AI adoption and provides a baseline for understanding how wider use of AI agents may further change the work clients purchase from freelancers.

Looking within AI-related Upwork job posts, \exref{task-composition} shows a related change in the
type of work clients request. We define ``AI-related job posts'' as postings that request AI skills
(e.g., developing AI app integration), require AI tools, or involve improving output generated by AI
(details of classification are described in \hyperref[app:methodology]{Appendix~\ref*{app:methodology}}).

During this early phase, AI-related demand was already shifting toward creative production and applied
business workflows. Tasks such as video production, social-media campaign management, marketing
automation, and AI API integration gained share, suggesting that AI was becoming embedded in producing
finished outputs and supporting business functions. By contrast, AI-generated written content, training
programs, cloud infrastructure, data and systems integration, and several writing- and data-related
activities lost share. Building chatbots and machine-learning models remained the largest cluster, but
its share declined from 21.8 percent to 17.8 percent and is excluded from the figure because of its
scale.

Most of the tasks that gained share within AI-related postings describe a finished product, such as
a video, or a business function, such as managing a marketing campaign. This suggests that clients
increasingly seek freelancers who can use AI as one part of delivering a broader project.

\begin{figure}[p]
\centering
\caption{Changes in the task mix of AI-related postings}\label{fig:task-composition}
\includegraphics[width=\linewidth,height=0.79\textheight,keepaspectratio]{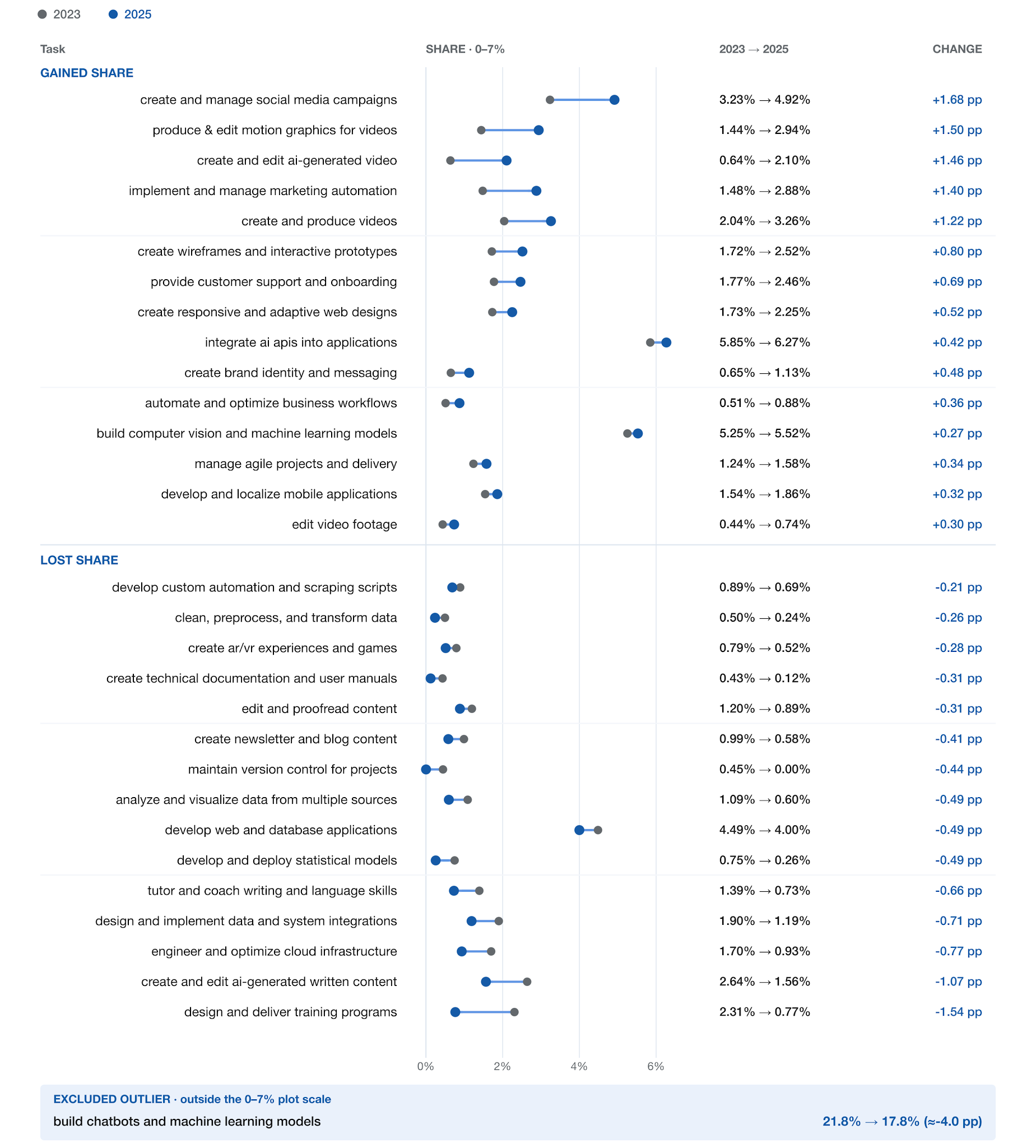}
\par\smallskip
\begin{minipage}{\linewidth}\footnotesize\raggedright
\textit{Note.} The figure shows the 15 task clusters with the largest share gains and the 15 with the largest share declines within AI-related work between 2023 and 2025. Shares are calculated as a percentage of all task-cluster assignments in AI-related job posts. Task clusters are not mutually exclusive, and a single AI-related job post may contain both a general production task, such as ``Create and produce videos,'' and a more specific AI-related task, such as ``Create and edit AI-generated video.'' AI-related classification applies to the job post as a whole, rather than requiring every task cluster within the post to be explicitly AI-specific. ``Build chatbots and machine-learning models'' was the largest task cluster in both years, but is not shown because its substantially larger scale would compress variation among the remaining clusters. Its share declined from 21.8\% in 2023 to 17.8\% in 2025.

\textit{Source.} OpenAI analysis of Upwork Marketplace data.

\end{minipage}
\end{figure}

\subsection{Price and quantity margins of AI-related work}

AI-related job posts on the Upwork Marketplace carry substantially higher posted project budgets than
non-AI postings. When requesting freelance work on Upwork, clients can choose between posting a
fixed-price budget for a project or an hourly rate they are willing to pay. We compare AI-related and non-AI postings separately for fixed-price projects and hourly work. Fixed-price budgets
capture the proposed payment for a project or deliverable, while posted hourly rates capture the proposed
price of worker time. For hourly postings that result in a hire, we also examine hours worked during the
first 90 days of the contract. Together, these measures help distinguish differences in project budgets,
the price per hour, and the amount of labor used.

\exref{prices-and-hours} shows that fixed-price budgets are approximately 70 percent higher for AI-related jobs overall and remain about 52
percent higher after controlling for work subcategory and posting month. The gap in posted hourly rates
is smaller, where the gap between AI and non-AI-related jobs is only about 9 percent and falls to roughly
1.5 percent after the same adjustment.

This contrast reflects an important difference between the two contract types. In a fixed-price contract,
the client purchases a project or deliverable as a whole, so the posted budget incorporates both the
value and the scope of the work. Hourly contracts allow us to distinguish between the posted price of
worker time and the quantity of labor purchased. While adjusted differences in posted hourly rates are
small, AI-related work involves substantially more hours per hired contract. Among hired hourly contracts
with positive recorded hours, hours worked during the first 90 days after hire are approximately 41
percent higher for AI-related work, controlling for work subcategory and calendar month. The AI/non-AI
gap is therefore much more pronounced in the amount of labor purchased per hired contract than in the
posted price per hour.

This pattern also helps interpret the large fixed-price gap. In a fixed-price contract, the amount of
labor, project scope, and value of the final deliverable are bundled into a single budget. Larger
AI-related budgets may therefore reflect larger or more intensive projects rather than simply a higher
price for an otherwise identical unit of labor. Taken together, the large fixed-price budget gap,
negligible hourly-rate gap, and substantially greater realized hours point to a common pattern:
AI-related work is associated with greater demand for labor and larger projects, rather than primarily
with a higher price for each hour of worker time. For example, AI may increase the returns to
experimentation and refinement, encouraging clients to explore more alternatives and invest additional
time in improving the quality of the final output, or to having workers complete and ``own'' larger, more
complex bundles of tasks. This mechanism is analogous to the rebound effect in the energy literature,
where lower costs can encourage greater use (\hyperlink{ref14}{Jevons, 1865}; \hyperlink{ref19}{Sorrell, 2009}).

This distinction between price and quantity margins is consistent with recent evidence that labor-market
adjustment associated with AI can operate strongly through employment and hiring while wage responses are
smaller or more heterogeneous (\hyperlink{ref6}{Chandar and Klein Teeselink, 2026};
\hyperlink{ref17}{Rinz, 2026}; \hyperlink{ref9}{de Souza, 2025}). The freelance setting makes this
distinction particularly transparent because hourly contracts separately reveal a requested price for
worker time and the subsequent quantity of labor used.

\begin{figure}[htbp]
\centering
\caption{Price and hours differences between AI-related and non-AI job posts}\label{fig:prices-and-hours}
\includegraphics[width=\linewidth,height=0.79\textheight,keepaspectratio]{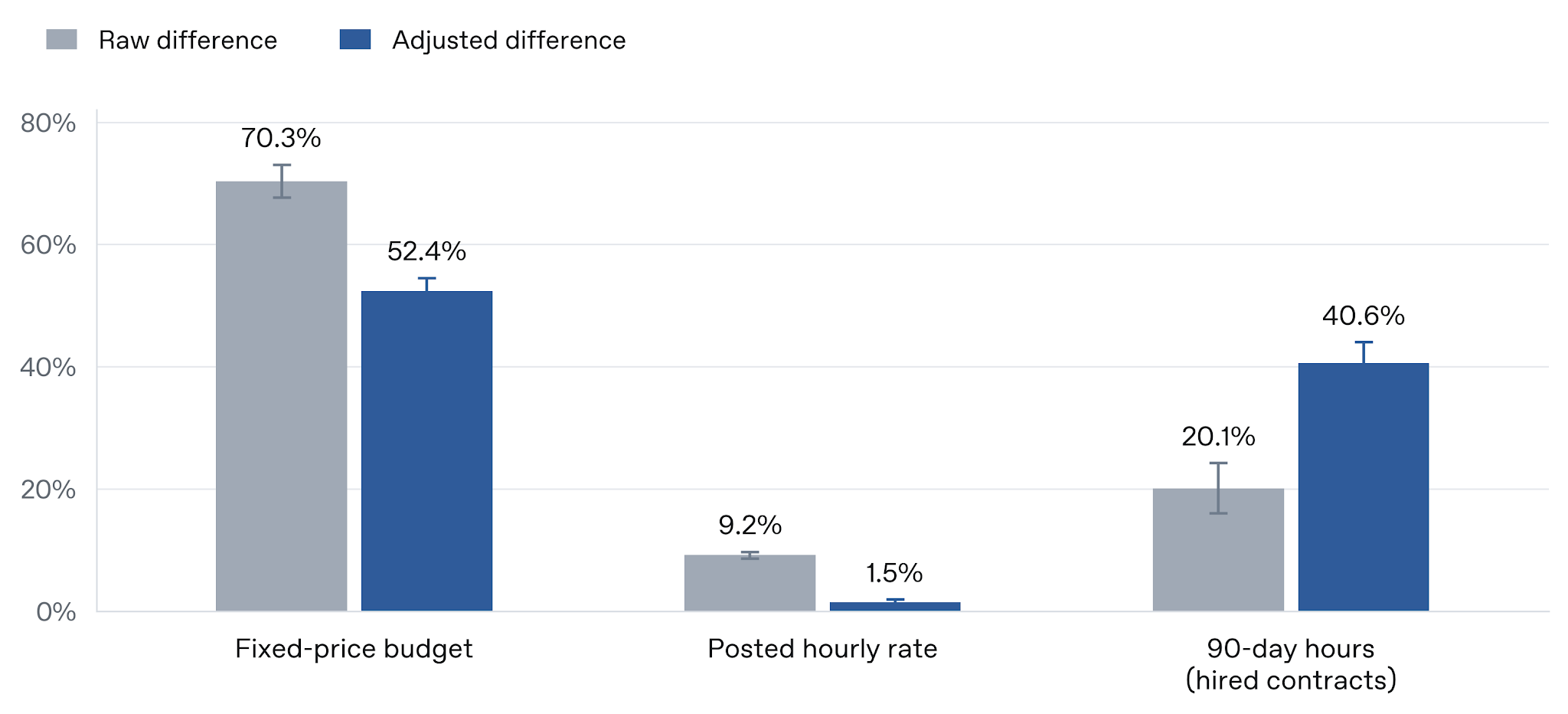}
\par\smallskip
\begin{minipage}{\linewidth}\footnotesize\raggedright
\textit{Note.} Bars show percentage differences between AI-related and non-AI job posts in fixed-price budgets, posted hourly rates, and hours worked in the first 90 days after hire on hourly contracts with positive recorded hours. Raw differences are unadjusted; adjusted estimates control for work subcategory and posting month. Error bars show 95\% confidence intervals. The top 1\% of posted prices is trimmed.

\textit{Source.} Upwork Marketplace data.

\end{minipage}
\end{figure}

\section{AI is becoming embedded in broader and more novel workflows}

As AI changes the relative costs of performing tasks and coordinating across tasks, it is likely to
affect both which tasks are performed and which tasks are performed together. We examine this possibility
in three steps. First, we ask whether AI-related demand is spreading across a broader range of tasks.
Second, we examine whether those tasks are being combined into broader projects. Finally, among projects with similar numbers of tasks, do AI-related postings combine tasks that
clients less often requested together before the generative AI era?

\subsection{AI-related work spans more activities and larger bundles}

We summarize the overall task mix across AI-related job postings using the effective number of task
clusters. A higher value means that work is less concentrated in a few activities. Between early 2023 and
late 2025, the effective number of task clusters increased from 14.6 to 21.2---approximately 45 percent.
An effective count of 21.2 tasks means that its concentration is equivalent to a distribution with about
21 equally represented clusters.

Thus, AI-related work is becoming less concentrated in a few dominant activities and more evenly spread
across activities. This broadening is consistent with AI becoming embedded in a wider range of economic
activities rather than remaining concentrated in a small number of early applications such as chatbot
development or text generation. We plot the effective number of task clusters in AI-related job postings
for comparison across these two periods in Panel A of \exref{task-diversity}.

\begin{figure}[htbp]
\centering
\caption{Task diversity in job postings}\label{fig:task-diversity}
\includegraphics[width=\linewidth,height=0.79\textheight,keepaspectratio]{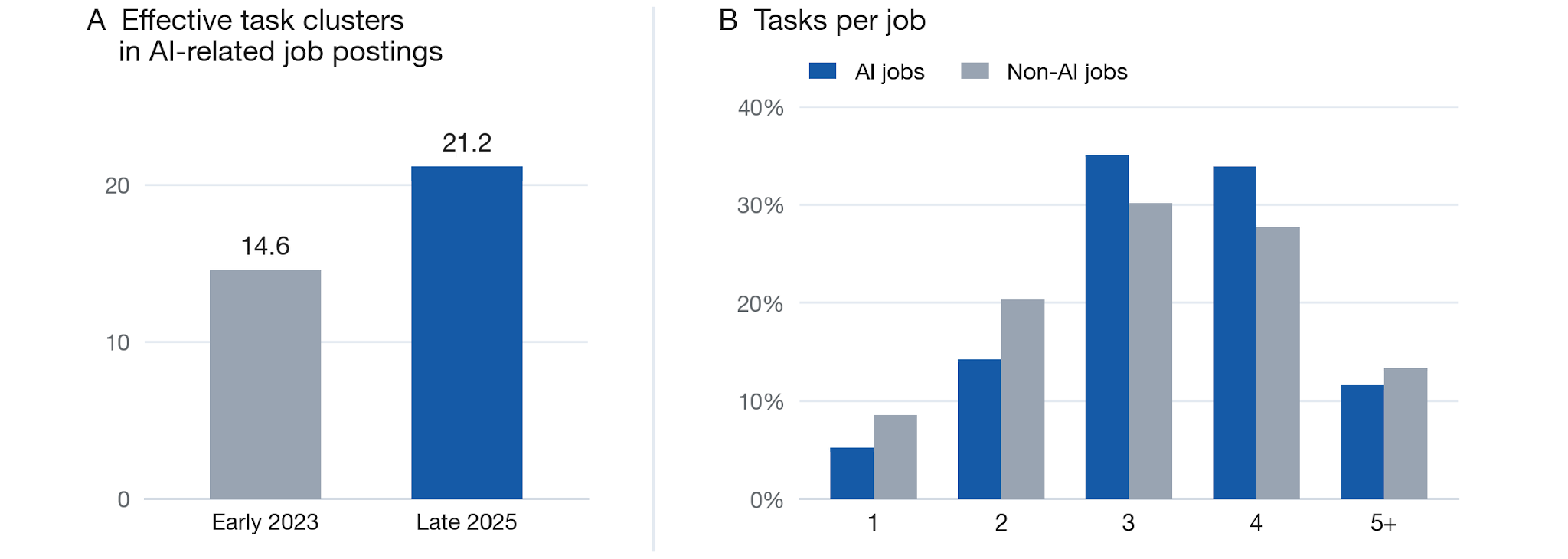}
\par\smallskip
\begin{minipage}{\linewidth}\footnotesize\raggedright
\textit{Note.} Data cover Upwork Marketplace job postings through 2025. Panel A shows the effective number of task clusters ($1/\mathrm{HHI}$) across AI-related task assignments; early/late average Q1--Q2 2023 and Q3--Q4 2025. Panel B shows the share of AI and non-AI posts by distinct task-cluster count. Means per post: 3.35 AI, 3.19 non-AI.

\end{minipage}
\end{figure}

Having shown that AI-related work is distributed across a wider range of economic activities, we next
examine whether this breadth also appears within individual jobs. To do so, we compare the number of
distinct task clusters in AI-related and non-AI-related job postings.

Panel B of \exref{task-diversity} plots the distribution of the number of identified tasks per job for AI-related and non-AI-related jobs. AI-related jobs contain a slightly larger number of distinct tasks on average; AI-related jobs include 3.35 distinct tasks per post, compared with 3.19 for non-AI jobs---a difference of approximately 5 percent. Although the average difference is modest, the distribution shows that it is not driven by a small number of unusually large AI projects. Instead, AI jobs are less likely to involve one or two distinct tasks, and more likely to involve three or four, with no corresponding increase in the five-or-more category. Therefore, AI-related work appears to take the form of more complex, multi-component projects, rather than being disproportionately associated with the very largest task bundles.

As an illustration, some AI-related postings combine chatbot or machine-learning development with
computer vision and AI API integration, while others combine AI-generated video with video production,
editing, and motion graphics. These projects bring AI-specific capabilities together with the
implementation and production tasks required to turn them into finished products.

\subsection{AI jobs combine tasks that were rarely paired before}

We next turn to the composition of the bundle itself. If AI is associated with a reorganization of work,
we might also expect to see changes in which tasks clients choose to purchase together. To understand
this, we compare how frequently AI-related and non-AI-related jobs combine tasks that were uncommon
together before the generative-AI era. For each current job posting containing at least two distinct
tasks, we identify all unique task pairs and count how many pre-AI postings contained each pair.

We classify pairs appearing in fewer than 200 pre-AI postings, including those with no observed
occurrence, as historically uncommon; pairs appearing in at least 200 pre-AI postings are established. We
then calculate the share of each job's task pairs meeting this definition and average those shares
separately for AI-related and non-AI-related jobs. Each job receives equal weight, regardless of how many
task pairs it contains.

The composition of these projects also differs. Historically uncommon combinations---task pairs appearing
in fewer than 200 pre-AI postings---account for an average of 8.4 percent of pairs within AI-related
jobs, compared with 2.4 percent within other jobs. This measure captures both combinations involving new
or historically rare tasks and combinations of familiar tasks that were seldom requested together. Panel
A of \exref{uncommon-pairs} plots this comparison. See \hyperref[tab:pair-composition]{Table~\ref*{tab:pair-composition}} for a more detailed
discussion.

\begin{figure}[htbp]
\centering
\caption{AI-related jobs and historically uncommon task pairings}\label{fig:uncommon-pairs}
\includegraphics[width=\linewidth,height=0.79\textheight,keepaspectratio]{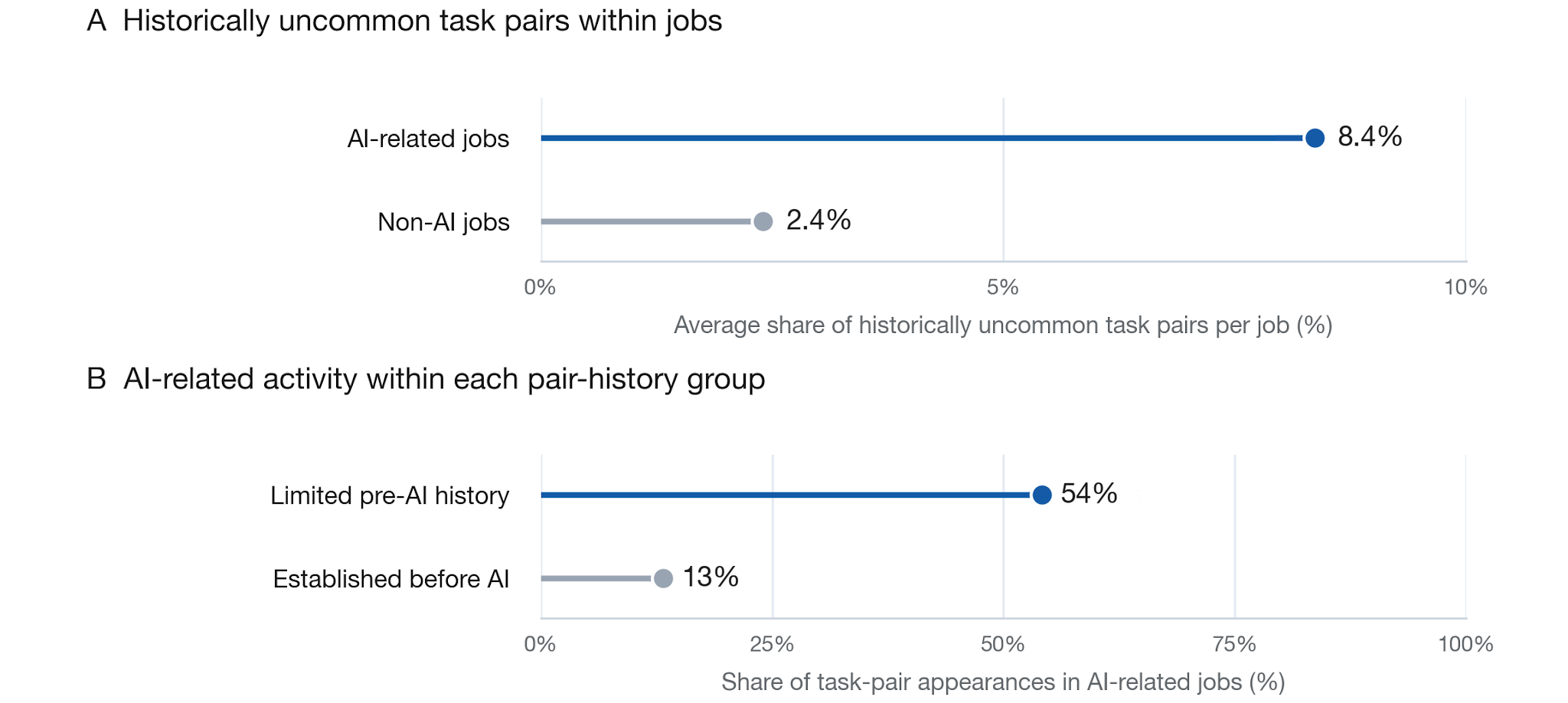}
\par\smallskip
\begin{minipage}{\linewidth}\footnotesize\raggedright
\textit{Note.} Data cover Upwork Marketplace job postings through 2025. Panel A shows the average within-job share of historically uncommon task pairs, separately for AI-related and non-AI jobs. Panel B groups task pairs appearing in at least 200 current job postings by their pre-AI history: limited history (fewer than 200 pre-AI postings, including zero) and established (at least 200). For each group, Panel B shows the share of current pair appearances occurring in AI-related jobs, weighted by the frequency of each pair.

\end{minipage}
\end{figure}

Panel B offers a complementary perspective. Among task pairs appearing in at least 200 current postings,
AI-related jobs account for 54 percent of appearances of historically uncommon pairs, compared with 13
percent for established pairs.

\begin{figure}[!t]
\centering
\caption{Task-bundle novelty by bundle size and type of AI work}\label{fig:bundle-novelty}
\includegraphics[width=0.92\linewidth,height=0.79\textheight,keepaspectratio]{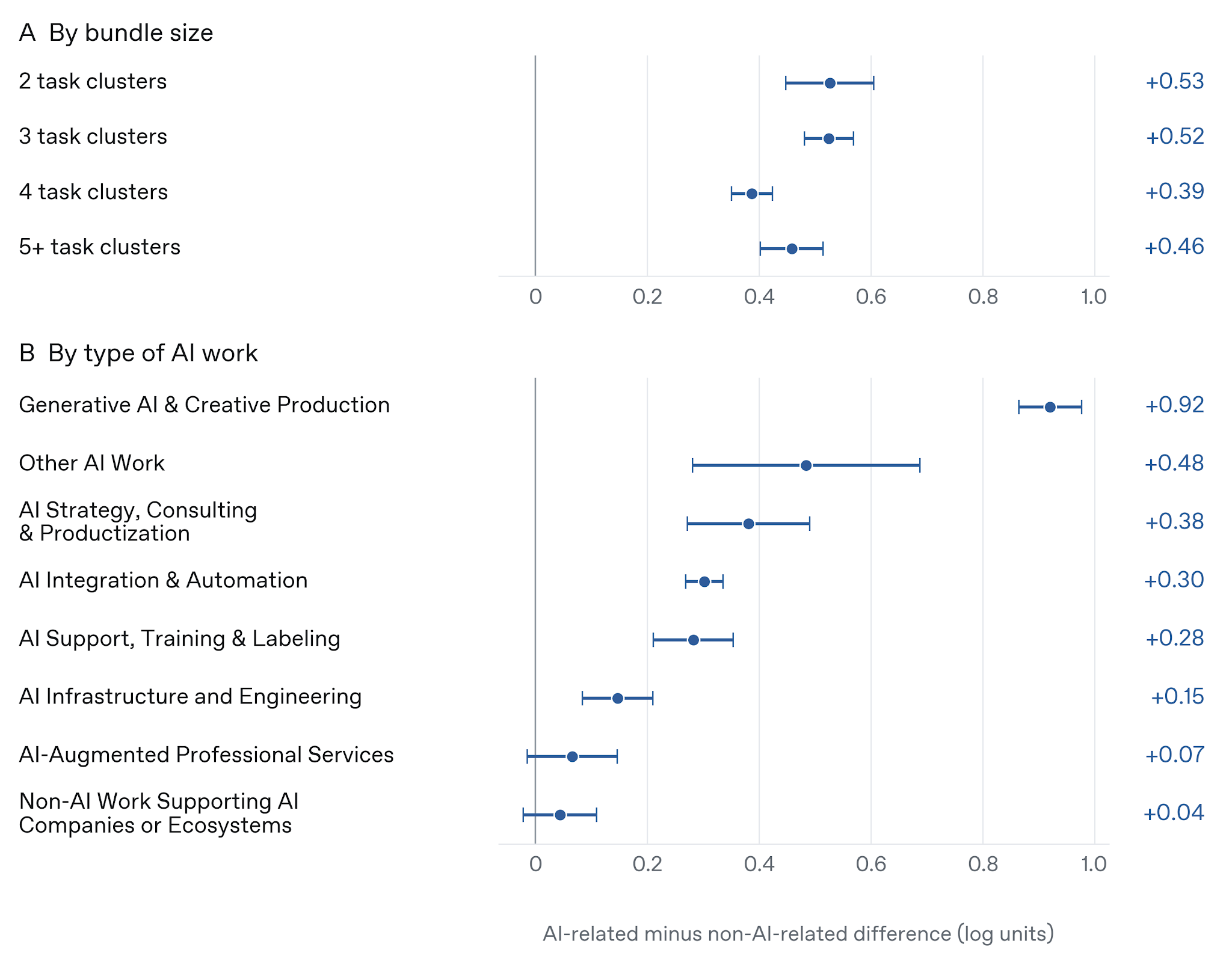}
\par\smallskip
\begin{minipage}{\linewidth}\footnotesize\raggedright
AI jobs combine tasks in historically less common ways. Panel A compares AI-related and non-AI-related
jobs within four bundle-size groups; Panel B compares eight types of AI work with non-AI-related jobs.
Positive coefficients indicate lower pre-AI prevalence of task pairings.

\textit{Note.} Data cover Upwork Marketplace job postings through 2025. The outcome is negative mean log pre-AI pair prevalence, using pairs observed in at least 20 pre-AI posts; posts without eligible pairs are excluded. Points show conditional differences from separate OLS models controlling for bundle-size group, work subcategory, and posting month. Both models use 104,030 posts from a 5\% sample. Bars show 95\% confidence intervals using HC1 standard errors.

\textit{Source.} OpenAI analysis of Upwork Marketplace data.

\end{minipage}
\end{figure}

\exref{bundle-novelty} focuses on task pairs observed in at least 20 pre-AI postings. This allows us to compare historical prevalence among previously observed pairings.

For each job posting, we calculate the negative natural logarithm of each task pair's historical
prevalence and average across eligible pairs within each job posting. The minus sign gives rarer pairings
higher scores, and averaging these scores across eligible pairs gives each job one overall novelty score.
The logarithmic scale compares proportional differences in frequency: the novelty gap between pairings
found in 10 percent and 1 percent of earlier postings is the same as the gap between pairings found in 1
percent and 0.1 percent. We calculate the measure using task pairs observed in at least 20 pre-AI
postings, allowing us to compare how common or rare combinations were among pairings already observed
before the generative-AI era. Entirely new pairings, including those involving new tasks, and other pairs
below this threshold are excluded from the score. We then compare AI-related and non-AI-related postings
while accounting for bundle size, work subcategory, and posting month. Because the measure is expressed
in logs, a coefficient of 0.5 corresponds to approximately 39 percent lower geometric mean historical
prevalence of task pairings.

After accounting for bundle size, work subcategory, and posting month, AI-related jobs still contain
combinations with lower historical prevalence, as shown by Panel A of \exref{bundle-novelty}. The
pattern holds across every bundle size examined. Among jobs containing two or three tasks, their pairings
had roughly 40 percent lower geometric mean historical prevalence. The differences are smaller for larger
bundles but remain statistically significant.

These controls allow the individual tasks to differ across jobs, so the comparison captures differences
in both the tasks selected and their combinations. We therefore interpret the result as evidence that
AI-related projects draw on bundles with less historical precedent.

The difference is particularly pronounced in generative AI and creative production, as shown by Panel B
of \exref{bundle-novelty}. Their task pairings had approximately 60 percent lower geometric mean
historical prevalence than those in comparable non-AI jobs.

AI strategy and productization, as well as AI integration and automation, also exhibit statistically
significant novelty, although the differences are smaller. AI support, training, and labeling and AI
infrastructure and engineering also show positive, statistically significant differences on this measure.
AI-augmented professional services and work supporting AI companies remain relatively close to the non-AI
baseline. Differences in the historical familiarity of task combinations are therefore concentrated in
particular areas, most notably generative AI and creative production.

Taken together, the results suggest that AI is becoming embedded in a wider range of freelance projects,
often alongside traditional implementation and production work. AI-generated video combined with editing
and motion graphics, or chatbot development combined with application integration, illustrates how
clients purchase capabilities together to obtain a finished outcome. The value of freelance work may
increasingly lie in connecting these capabilities, applying judgment, and completing the surrounding
workflow. These patterns provide an early baseline for understanding how more agentic AI may further
change the activities clients purchase together.

\section{Conclusion}

The findings suggest that AI is changing both the division of work and the organization of freelance
projects. ChatGPT use is relatively concentrated in writing and advice, while Upwork postings place
greater emphasis on technical implementation, visual production, and specialized delivery. Within
AI-related freelance work, demand is spreading across more activities and shifting toward creative
production and applied business workflows. AI-related jobs also combine tasks in less historically
familiar ways, even among projects of comparable breadth.

The contrast between larger project budgets, more hours per hired hourly contract, and a small difference
in posted hourly rates suggests that project scope is an important part of this adjustment. AI may make
individual activities easier to perform while encouraging clients to undertake more ambitious projects,
explore additional alternatives, or invest more effort in refinement. Projects that combine AI-generated
video with editing and motion graphics, or chatbot development with application integration, illustrate
how new capabilities can create demand for complementary work.

These patterns connect the adoption of AI to the economics of organizing work. By changing the costs of
performing activities and coordinating across them, AI may alter which tasks clients find useful to
purchase together. Value may increasingly lie in defining the problem, connecting tools and activities,
applying domain judgment, and delivering a finished outcome. The resulting adjustment can redraw project
boundaries as clients assemble different combinations of capabilities to achieve their goals.

Upwork and ChatGPT do not represent the labor market as a whole, but together they offer an early view of
a broader shift: AI is changing not just the tasks people perform, but the way those tasks are divided,
combined, and brought to market.

\clearpage\phantomsection\addcontentsline{toc}{section}{References}

\section*{References}\begingroup\linespread{1.0}\selectfont

\begin{list}{}{\setlength{\leftmargin}{1.5em}\setlength{\itemindent}{-1.5em}\setlength{\itemsep}{1.2pt}\setlength{\parsep}{0pt}}

\small

\item[] \hypertarget{ref1}{}Acemoglu, D., \& Restrepo, P. (2019). Automation and new tasks: How technology displaces and reinstates labor. Journal of Economic Perspectives, 33(2), 3--30. \url{https://doi.org/10.1257/jep.33.2.3}

\item[] \hypertarget{ref2}{}Autor, D. H. (2015). Why are there still so many jobs? The history and future of workplace automation. Journal of Economic Perspectives, 29(3), 3--30. \url{https://doi.org/10.1257/jep.29.3.3}

\item[] \hypertarget{ref3}{}Autor, D., Chin, C., Salomons, A., \& Seegmiller, B. (2024). New frontiers: The origins and content of new work, 1940--2018. The Quarterly Journal of Economics, 139(3), 1399--1465. \url{https://doi.org/10.1093/qje/qjae008}

\item[] \hypertarget{ref4}{}Becker, G. S., \& Murphy, K. M. (1992). The division of labor, coordination costs, and knowledge. Quarterly Journal of Economics, 107(4), 1137--1160. \url{https://doi.org/10.2307/2118383}

\item[] \hypertarget{ref5}{}Brynjolfsson, E., Li, D., \& Raymond, L. R. (2025). Generative AI at work. Quarterly Journal of Economics, 140(2), 889--942. \url{https://doi.org/10.1093/qje/qjae044}

\item[] \hypertarget{ref6}{}Chandar, B., \& Klein Teeselink, B. (2026). How does AI change labor demand? Evidence from 41 countries. Stanford Digital Economy Lab. \url{https://digitaleconomy.stanford.edu/publication/how-does-ai-change-labor-demand/}

\item[] \hypertarget{ref7}{}Dell'Acqua, F., McFowland, E., III, Mollick, E., Lifshitz, H., Kellogg, K. C., Rajendran, S., Krayer, L., Candelon, F., \& Lakhani, K. R. (2026). Navigating the jagged technological frontier: Field experimental evidence of the effects of artificial intelligence on knowledge worker productivity and quality. Organization Science, 37(2), 403--423. \url{https://doi.org/10.1287/orsc.2025.21838}

\item[] \hypertarget{ref8}{}Demirer, M., Horton, J. J., Immorlica, N., Lucier, B., \& Shahidi, P. (2026). Chaining tasks, redefining work: A theory of AI automation (NBER Working Paper No. 34859). National Bureau of Economic Research. \url{https://www.nber.org/papers/w34859}

\item[] \hypertarget{ref9}{}de Souza, G. (2025). AI in the office and the factory: Evidence from administrative software registry data (Working Paper No. 2025-11, revised July 2026). Federal Reserve Bank of Chicago. \url{https://doi.org/10.21033/wp-2025-11}

\item[] \hypertarget{ref10}{}Eloundou, T., Manning, S., Mishkin, P., \& Rock, D. (2024). GPTs are GPTs: Labor market impact potential of LLMs. Science, 384(6702), 1306--1308. \url{https://doi.org/10.1126/science.adj0998}

\item[] \hypertarget{ref11}{}Gans, J. S. (2026). Endogenous Task Bundling, Skills and Automation (Working Paper No. 35211). National Bureau of Economic Research. \url{https://doi.org/10.3386/w35211}

\item[] \hypertarget{ref12}{}Garicano, L., Li, J., \& Wu, Y. (2026). Weak bundle, strong bundle: How AI redraws job boundaries (CEPR Discussion Paper No. DP21453). Centre for Economic Policy Research. \url{https://cepr.org/publications/dp21453}

\item[] \hypertarget{ref13}{}Hui, X., Reshef, O., \& Zhou, L. (2024). The short-term effects of generative artificial intelligence on employment: Evidence from an online labor market. Organization Science, 35(6), 1977--1989. \url{https://doi.org/10.1287/orsc.2023.18441}

\item[] \hypertarget{ref14}{}Jevons, W. S. (1865). The Coal Question: An Inquiry Concerning the Progress of the Nation, and the Probable Exhaustion of Our Coal-Mines. Macmillan.

\item[] \hypertarget{ref15}{}Noy, S., \& Zhang, W. (2023). Experimental evidence on the productivity effects of generative artificial intelligence. Science, 381(6654), 187--192. \url{https://doi.org/10.1126/science.adh2586}

\item[] \hypertarget{ref16}{}Ocampo, S. (2022). A task-based theory of occupations with multidimensional heterogeneity (Working Paper No. 20222). Western University, Department of Economics. \url{https://ideas.repec.org/p/uwo/hcuwoc/20222.html}

\item[] \hypertarget{ref17}{}Rinz, K. (2026). The recent evolution of AI-related labor demand (Working Paper No. 26-24). Federal Reserve Bank of Cleveland. \url{https://doi.org/10.26509/frbc-wp-202624}

\item[] \hypertarget{ref18}{}Rosen, S. (1978). Substitution and division of labour. Economica, 45(179), 235--250. \url{https://doi.org/10.2307/2553070}

\item[] \hypertarget{ref19}{}Sorrell, S. (2009). Jevons' paradox revisited: The evidence for backfire from improved energy efficiency. Energy Policy, 37(4), 1456--1469. \url{https://doi.org/10.1016/j.enpol.2008.12.003}

\item[] \hypertarget{ref20}{}Yiu, S., Seamans, R., Raj, M., \& Liu, T. (2026). Worker repositioning and technological change: Evidence from an online labor market. Forthcoming in Organization Science. \url{https://doi.org/10.1287/orsc.2025.20566}

\end{list}\endgroup

\clearpage\appendix

\section{Composition of task-cluster pairings}

\exref{uncommon-pairs} shows that AI-related jobs contain a greater share of task pairs with limited pre-AI precedent. This section examines the composition of historically uncommon pairings within AI-related jobs. In particular, we ask whether historically uncommon pairs in AI-related jobs primarily join AI-associated tasks with traditional work, or instead reflect traditional tasks being combined with one another in less familiar ways.

Most historically uncommon pair occurrences in current AI-related jobs consist of two tasks that each
appeared frequently in pre-period postings. Of such pair occurrences, 98.6 percent involve two tasks that were
each observed in at least 200 pre-period postings. Thus, historically uncommon combinations do not
generally consist of two wholly new tasks. Rather, they often join tasks that each had an observed
pre-period history but were seldom requested together.

To further characterize these combinations, we apply a curated AI-task definition at the task level. The
curated list maps 141 normalized inferred task labels to eight AI-focused work specializations: AI
Chatbot Development; AI Content Writing; AI Data Annotation \& Labeling; AI Image Generation \& Editing;
AI Integration; AI Speech \& Audio Generation; AI Video Generation \& Editing; and Generative AI
Modeling. A task cluster is classified as AI-associated if at least half of its component task labels
appear in this dictionary; otherwise, it is classified as traditional.

The benefit of this approach is that it classifies clusters according to their task content rather than
according to how often they appear in jobs already labeled AI-related. It therefore reduces the
mechanical overlap between the historically uncommon pairings being described and the rule used to
identify an AI-associated cluster. If the same mixed AI--traditional pattern appears under both the
curated and the report's existing usage-based measure of task-level AI exposure, the interpretation does
not depend solely on defining a task as AI-intensive because it is frequently used in AI-related jobs.

For completeness, we also use the report's existing job post measure of task-level AI exposure: the share
of current postings containing a task cluster that are classified as AI-related. For this decomposition,
we classify clusters with an AI-related share of at least 20 percent as AI-intensive and all other clusters as
traditional. This converts the report's continuous exposure measure into a binary cluster classification
for the purpose of categorizing task pairs.

Using each of the two definitions, we classify task pairs as AI-only (both clusters are AI-associated or AI-intensive), mixed (one cluster is
AI-associated or AI-intensive and the other is traditional), or fully traditional (neither cluster meets
the AI rule).

\begin{table}[htbp]
\centering
\caption{Composition of historically uncommon task-cluster pairings within AI-related jobs}\label{tab:pair-composition}
\small
\setlength{\tabcolsep}{4pt}
\renewcommand{\arraystretch}{1.17}
\begin{tabularx}{\linewidth}{@{} l l >{\raggedleft\arraybackslash}X >{\raggedleft\arraybackslash}X >{\raggedleft\arraybackslash}X @{}}
\toprule
Definition & Pair type & Pairs per 100 total & Historically uncommon rate & Uncommon pairs per 100 total \\
\midrule
Curated & Fully traditional & 60.1 & 3.4\% & 2.0 \\
 & Mixed AI--traditional & 32.8 & 12.1\% & 4.0 \\
 & AI-only & 7.2 & 15.7\% & 1.1 \\
 & All pair types & 100.0 & --- & 7.1 \\
\midrule
Usage-based & Fully traditional & 49.8 & 2.2\% & 1.1 \\
 & Mixed AI--traditional & 26.8 & 16.0\% & 4.3 \\
 & AI-only & 23.4 & 7.3\% & 1.7 \\
 & All pair types & 100.0 & --- & 7.1 \\
\bottomrule
\end{tabularx}
\par\smallskip
\begin{minipage}{\linewidth}\footnotesize\raggedright
\textit{Note.} Results use the report's one-in-20 sample of current AI-related jobs. For each job, we enumerate all distinct within-job task-cluster pairs; each job--pair occurrence is counted as one observation. We apply both the curated and usage-based cluster classifications to this same set of job--pair observations. The overall historically uncommon-pair rate in this table does not equal the 8.4\% reported in Panel A of \exref{uncommon-pairs} because the two summaries use different weights. \exref{uncommon-pairs} gives each job equal weight by first calculating each job's uncommon-pair share and then averaging across jobs. This table pools job--pair observations, so jobs containing more task pairs receive greater weight.
\end{minipage}
\end{table}

Under both definitions, mixed AI--traditional pairs are the largest component of historically uncommon
pair occurrences in AI-related jobs: about four per 100 total pair occurrences, compared with roughly one
to two for each other pair type. This pattern is consistent with AI-related jobs bringing AI-associated
capabilities together with traditional work in combinations that had limited historical precedent. It
should not be interpreted as showing that AI has broadly caused traditional tasks to be recombined with
one another.

\section{Methodology}\label{app:methodology}

\methodhead{AI-related work classification}

We use Upwork's internal AI-related jobs pipeline (a GPT-powered LLM with structured prompting) designed
to identify AI-related jobs on the platform. The model evaluates the full context of each job post,
including the job title, description, requested skills, category/subcategory/specialization ontology, and
other job characteristics and information, when applicable, to determine whether the job is AI-related.
As AI evolves rapidly, this context-aware approach identifies AI-related jobs beyond predefined keywords
by understanding broader context and client intent. The pipeline runs at scale across job posts and
produces a standardized AI-related job classification for downstream analytics and reporting.

\methodhead{Task construction}\hypertarget{task-construction}{}

We use detailed activity descriptions to represent the types of work performed on Upwork. We create a
library of 3,000+ activity descriptions from Upwork's occupational taxonomy, and historical job posts. We
then group similar activity descriptions into 400+ standardized tasks, also called task clusters in the
technical methodology, using embedding-based clustering, using UMR (Upwork's Meaningful Representation),
Upwork's semantic AI model, with additional AI-assisted and manual review to improve consistency and
interpretability. This creates a stable framework for tracking how different types of work, job volume,
and value change over time, including shifts associated with the adoption of AI.

\methodex{activity-shares}

\exref{activity-shares} presents analyses of sampled ChatGPT conversations and tasks posted in the Upwork Marketplace. The ChatGPT analysis in the left-hand panel uses conversations from individual consumer ChatGPT accounts sampled separately within each month from January 2025 through June 2026, targeting 5,556 conversations per month. Each conversation is classified using up to its first five available user messages and assigned a primary task from Upwork's 64-category task taxonomy, or labeled as having no matching task. The final sample includes 36,896 conversations classified as involving personal or work-related tasks. We exclude schoolwork, and conversations with unclear purpose, and conversations that are not linked to an Upwork task. Reported task shares are the proportion of all conversations in the final sample assigned to each task. All classifications were conducted using GPT 5.6 Luna; no researcher read user messages or looked at individual user data in the course of this research. \hyperref[app:chatgpt-prompt]{Appendix~\ref*{app:chatgpt-prompt}} reproduces the full classifier text.

For Upwork, we calculate each subcategory's share of all Marketplace postings in January through June
of 2025 and in the same six months of 2026. The sample includes AI-related and other postings. We
subtract the H1 2025 share from the H1 2026 share to obtain the change in percentage points. The
chart shows the 15 largest gains and the 15 largest declines, alongside each subcategory's H1 2026
share.

\methodex{chatgpt-intensity}

Each point represents one matched work subcategory in the Upwork taxonomy. The horizontal axis measures
ChatGPT intensity: the natural log of the category's share of sampled ChatGPT conversations in H2 2025
divided by its share of Upwork Marketplace job posts in H1 2023. Higher values indicate activities that
were relatively more prevalent in direct ChatGPT use than in the baseline freelance market.

The vertical axis measures relative category growth: the category's H2 2025-to-H1 2023 posting-growth
factor divided by the corresponding growth factor for total Marketplace postings. Equivalently, it is the
ratio of the category's H2 2025 Marketplace share to its H1 2023 Marketplace share, shown on a
logarithmic scale. Values above the 0 percent reference line indicate that a category gained Marketplace share;
values below it indicate that the category lost share.

The analysis includes 63 matched work subcategories, excluding AI Apps \& Integration and categories with
zero H1 2023 Upwork posts. The figure does not measure aggregate
Marketplace growth or identify a causal effect of ChatGPT use on Upwork demand.

\methodex{task-composition}

The figure compares the task-cluster mix of AI-related Upwork job posts in calendar years 2023 and 2025.
We assign each task in an AI-related job post to an inferred task cluster. For each year, we calculate
each cluster's share of all task-cluster assignments in AI-related work. The figure plots clusters whose
share was at least 0.25 percent in either year. Grey markers show the cluster's 2023 share, and colored markers
show its 2025 share; connecting lines show the change in share, in percentage points. For readability,
the display includes the 15 largest share gains and 15 largest share declines, excluding ``Build chatbots
and machine learning models'' because of its substantially larger scale. The figure describes changes in
the composition of AI-related work, not changes in aggregate Upwork Marketplace demand or total posting
volume.

\methodex{prices-and-hours}

We compare posted prices and hours worked in the first 90 days after hiring for AI-related and
non-AI-related job posts in the current period, through April 2026. The fixed-price outcome is the posted
project budget; the hourly-price outcome is the midpoint of the posted hourly-rate range; and the third
outcome is total hours worked in the first 90 days after hire for hired hourly contracts. Analyses use
positive values only and trim the top 1 percent of the relevant posted-price distribution.

Raw estimates compare arithmetic means for AI-related and non-AI-related posts. Adjusted estimates are
based on separate post-level OLS regressions of the log outcome on an AI-related-job indicator,
calendar-month fixed effects, and work-subcategory fixed effects, with heteroskedasticity-consistent
standard errors. The adjusted estimates are expressed as the percentage difference associated with
AI-related classification, relative to non-AI-related posts in the same month and work subcategory. For
the 90-day-hours outcome, the sample is limited to hired hourly posts with positive recorded hours, and
the adjustment uses the same month and work-subcategory controls. These comparisons are descriptive and
do not identify a causal effect of AI-related classification.

\methodex[ (Panel A)]{task-diversity}

We use inferred task assignments from Upwork job posts classified as AI-related between January 2023 and
December 2025. Each inferred task is mapped to an Upwork task cluster. For each calendar month, we count
assignments to each cluster and calculate that cluster's share of all AI-related task-cluster
assignments. We calculate the Herfindahl-Hirschman Index (HHI) as the sum of squared cluster shares and
report its inverse, effective $N$ ($1/\mathrm{HHI}$), which expresses concentration as the equivalent number of
equally represented task clusters. Higher values indicate a more dispersed mix of task-cluster
assignments.

\methodex[ (Panel B)]{task-diversity}

We compare the distribution of task-bundle size in current AI-related and non-AI-related job posts. For
each post, we count the number of distinct inferred task clusters associated with that post; repeated
tasks mapped to the same cluster are counted once. We then classify posts as containing 1, 2, 3, 4, or 5
or more distinct task clusters and calculate the share of posts in each category separately for
AI-related and non-AI-related work. Each post receives equal weight, and the 5-or-more category combines
all larger task bundles.

This panel is descriptive and shows that AI-related posts tend to contain somewhat broader bundles of distinct task clusters.

\methodex[ (Panel A)]{uncommon-pairs}

We compare the historical prevalence of task combinations in current AI-related and non-AI-related job
posts. For each job with at least two mapped task clusters, we construct all unordered pairs of distinct
clusters. A pair is classified as having limited pre-AI precedent if it appeared together in fewer than
200 posts during the pre-AI period, including pairs with no observed pre-AI occurrence. We calculate, for
each job, the share of its within-job cluster pairs with limited pre-AI precedent, and report the average
of this measure separately for AI-related and non-AI-related jobs. The sample contains $N = 104{,}066$ job posts.

\methodex[ (Panel B)]{uncommon-pairs}

We examine whether task combinations with limited pre-AI precedent are disproportionately associated with
AI-related work in the current period. The analysis includes task-cluster pairs appearing in at least 200
current posts. Pairs are classified as having limited pre-AI precedent if they appeared in fewer than 200
pre-AI posts and as established if they appeared in at least 200. For each group, we calculate the share
of the current pair--post occurrences that come from AI-related jobs. Results are weighted by current
pair--post occurrences, such that a pair appearing in more current jobs receives greater weight; a job
containing multiple qualifying pairs may therefore contribute more than once.

\clearpage\begingroup\small\linespread{1.00}\selectfont
\methodex{bundle-novelty}

\methodhead{Panel A. By bundle size}

Panel A reports coefficients from a post-level ordinary least squares regression estimated on a one-in-20
hash sample of current job posts classified as AI-related or non-AI-related. The sample is limited to
posts containing at least two distinct inferred task clusters. The final regression sample contains
104,030 posts. For observed-pair novelty ($Y_i$), we estimate:

\begin{equation}
Y_i = \alpha_{b(i)} + \gamma_{\ell(i)} + \lambda_{m(i)}
      + \sum_b \beta_b[\mathrm{AI\ job}_i \times \mathbf{1}\{\mathrm{Bundle\ size}_i=b\}] + \varepsilon_i .
\end{equation}

Here $\alpha_{b(i)}$, $\gamma_{\ell(i)}$, and $\lambda_{m(i)}$ are bundle-size, work-subcategory, and
calendar-month fixed effects. $\mathrm{AI\ job}_i$ equals 1 if post $i$ is AI-related and 0 otherwise.
The index $b$ denotes the 2-, 3-, 4-, and 5+-task-cluster bundle-size buckets. The coefficient $\beta_b$
is the AI-versus-non-AI difference in novelty for posts in bucket $b$, and $\varepsilon_i$ is the
residual error.

Because the AI indicator is interacted with every bundle-size bucket, $\beta_b$ is the conditional
difference in the outcome between AI-related and non-AI-related posts with bundle size $b$, holding work
subcategory and posting month constant. Bars show 95 percent confidence intervals based on
heteroskedasticity-consistent (HC1) standard errors. These estimates describe conditional associations
and do not identify causal effects.

\methodhead{Observed-pair novelty (both panels)}

For each pair of task clusters $(a,b)$, pre-AI pair prevalence is defined as the number of pre-AI
posts containing both clusters divided by the number of pre-AI posts containing at least one mapped
task cluster. For each current post, the outcome is the negative mean log prevalence across its
within-post cluster pairs with recorded pre-AI support:

\begin{equation}
Y_i = -\frac{1}{|P_i|} \sum_{(a,b)\in P_i}\ln p_{ab},
\end{equation}
where $P_i$ is the set of qualifying within-post task-cluster pairs and $p_{ab}$ is their pre-AI pair prevalence.

Higher values indicate that the post's task pairs were less prevalent before the AI boom. In the source
data used here, the pre-AI pair table retains pairs observed in at least 20 pre-AI posts; pairs observed
fewer than 20 times are not included. Posts with no qualifying observed pre-AI pair are omitted from both
panels' regressions. A coefficient $\beta$ corresponds to $100\times(1-\exp(-\beta))$ percent lower
geometric mean historical pair prevalence.

\methodhead{Panel B. By type of AI work}

Panel B reports estimates from a post-level ordinary least squares regression using the same sample and observed-pair novelty outcome as Panel A. We estimate:

\begin{equation}
Y_i = \alpha + \sum_g \beta_g\,\mathbf{1}\{\mathrm{AI\ work\ group}_i=g\}
      + \gamma_{b(i)} + \delta_{\ell(i)} + \lambda_{m(i)} + \varepsilon_i .
\end{equation}

Here $i$ indexes job posts and $g$ indexes AI work categories. The indicator $\mathbf{1}\{\mathrm{AI\
work\ group}_i=g\}$ equals 1 when post $i$ belongs to category $g$ and 0 otherwise. The terms
$\gamma_{b(i)}$, $\delta_{\ell(i)}$, and $\lambda_{m(i)}$ represent bundle-size, work-subcategory, and
calendar-month fixed effects. Non-AI-related posts are the reference group.

For each AI work category, $\beta_g$ is the conditional difference in the novelty outcome relative to
non-AI-related posts, holding the bundle-size group, work subcategory, and posting month constant. Thus,
unlike Panel A, the model does not allow the AI-category gap to vary by bundle size.

The outcome is observed-pair novelty (negative mean log pre-AI pair prevalence), defined above. Bars show
95 percent confidence intervals based on heteroskedasticity-consistent (HC1) standard errors. The estimates
describe conditional associations and do not identify causal effects.

\endgroup
\clearpage

\section{Classifier text for ChatGPT data}\label{app:chatgpt-prompt}

The prompt below was used to classify the ChatGPT conversations analyzed in \exref{activity-shares}.

\begin{classifierprompt}
You are a classifier that helps assign conversations to a set of classified outputs.

Your job is to classify a conversation\textquotesingle{}s first user messages into one of the following categories of
work. In addition, classify whether the conversation pertains to work, schoolwork, personal or an unclear
purpose.

\methodhead{Tasks}

Below is a list of categories you will assign the message to.

\begingroup\footnotesize\setlength{\parskip}{0pt}\setlength{\columnsep}{18pt}\setlength{\parindent}{0pt}\begin{multicols}{2}\raggedcolumns

{\hangindent=4.35em\noindent\makebox[4.35em][l]{\texttt{U.1.1}}Accounting \& Bookkeeping\par}

{\hangindent=4.35em\noindent\makebox[4.35em][l]{\texttt{U.1.2}}Financial Planning\par}

{\hangindent=4.35em\noindent\makebox[4.35em][l]{\texttt{U.1.3}}Management Consulting \& Analysis\par}

{\hangindent=4.35em\noindent\makebox[4.35em][l]{\texttt{U.1.4}}Other - Accounting \& Consulting\par}

{\hangindent=4.35em\noindent\makebox[4.35em][l]{\texttt{U.1.5}}Personal \& Professional Coaching\par}

{\hangindent=4.35em\noindent\makebox[4.35em][l]{\texttt{U.1.6}}Recruiting \& Human Resources\par}

{\hangindent=4.35em\noindent\makebox[4.35em][l]{\texttt{U.2.1}}Data Entry \& Transcription Services\par}

{\hangindent=4.35em\noindent\makebox[4.35em][l]{\texttt{U.2.2}}Market Research \& Product Reviews\par}

{\hangindent=4.35em\noindent\makebox[4.35em][l]{\texttt{U.2.3}}Project Management\par}

{\hangindent=4.35em\noindent\makebox[4.35em][l]{\texttt{U.2.4}}Virtual Assistance\par}

{\hangindent=4.35em\noindent\makebox[4.35em][l]{\texttt{U.3.1}}Community Management \& Tagging\par}

{\hangindent=4.35em\noindent\makebox[4.35em][l]{\texttt{U.3.2}}Customer Service \& Tech Support\par}

{\hangindent=4.35em\noindent\makebox[4.35em][l]{\texttt{U.4.1}}AI \& Machine Learning\par}

{\hangindent=4.35em\noindent\makebox[4.35em][l]{\texttt{U.4.2}}Data Analysis \& Testing\par}

{\hangindent=4.35em\noindent\makebox[4.35em][l]{\texttt{U.4.3}}Data Extraction/ETL\par}

{\hangindent=4.35em\noindent\makebox[4.35em][l]{\texttt{U.4.4}}Data Mining \& Management\par}

{\hangindent=4.35em\noindent\makebox[4.35em][l]{\texttt{U.5.1}}Art \& Illustration\par}

{\hangindent=4.35em\noindent\makebox[4.35em][l]{\texttt{U.5.2}}Audio \& Music Production\par}

{\hangindent=4.35em\noindent\makebox[4.35em][l]{\texttt{U.5.3}}Branding \& Logo Design\par}

{\hangindent=4.35em\noindent\makebox[4.35em][l]{\texttt{U.5.4}}Graphic, Editorial \& Presentation Design\par}

{\hangindent=4.35em\noindent\makebox[4.35em][l]{\texttt{U.5.5}}NFT, AR/VR \& Game Art\par}

{\hangindent=4.35em\noindent\makebox[4.35em][l]{\texttt{U.5.6}}Performing Arts\par}

{\hangindent=4.35em\noindent\makebox[4.35em][l]{\texttt{U.5.7}}Photography\par}

{\hangindent=4.35em\noindent\makebox[4.35em][l]{\texttt{U.5.8}}Product Design\par}

{\hangindent=4.35em\noindent\makebox[4.35em][l]{\texttt{U.5.9}}Video \& Animation\par}

{\hangindent=4.35em\noindent\makebox[4.35em][l]{\texttt{U.6.1}}3D Modeling \& CAD\par}

{\hangindent=4.35em\noindent\makebox[4.35em][l]{\texttt{U.6.2}}Building \& Landscape Architecture\par}

{\hangindent=4.35em\noindent\makebox[4.35em][l]{\texttt{U.6.3}}Chemical Engineering\par}

{\hangindent=4.35em\noindent\makebox[4.35em][l]{\texttt{U.6.4}}Civil \& Structural Engineering\par}

{\hangindent=4.35em\noindent\makebox[4.35em][l]{\texttt{U.6.5}}Contract Manufacturing\par}

{\hangindent=4.35em\noindent\makebox[4.35em][l]{\texttt{U.6.6}}Electrical \& Electronic Engineering\par}

{\hangindent=4.35em\noindent\makebox[4.35em][l]{\texttt{U.6.7}}Energy \& Mechanical Engineering\par}

{\hangindent=4.35em\noindent\makebox[4.35em][l]{\texttt{U.6.8}}Interior \& Trade Show Design\par}

{\hangindent=4.35em\noindent\makebox[4.35em][l]{\texttt{U.6.9}}Physical Sciences\par}

{\hangindent=4.35em\noindent\makebox[4.35em][l]{\texttt{U.7.1}}Database Management \& Administration\par}

{\hangindent=4.35em\noindent\makebox[4.35em][l]{\texttt{U.7.2}}DevOps \& Solution Architecture\par}

{\hangindent=4.35em\noindent\makebox[4.35em][l]{\texttt{U.7.3}}ERP/CRM Software\par}

{\hangindent=4.35em\noindent\makebox[4.35em][l]{\texttt{U.7.4}}Information Security \& Compliance\par}

{\hangindent=4.35em\noindent\makebox[4.35em][l]{\texttt{U.7.5}}Network \& System Administration\par}

{\hangindent=4.35em\noindent\makebox[4.35em][l]{\texttt{U.8.1}}Corporate \& Contract Law\par}

{\hangindent=4.35em\noindent\makebox[4.35em][l]{\texttt{U.8.2}}Finance \& Tax Law\par}

{\hangindent=4.35em\noindent\makebox[4.35em][l]{\texttt{U.8.3}}International \& Immigration Law\par}

{\hangindent=4.35em\noindent\makebox[4.35em][l]{\texttt{U.8.4}}Public Law\par}

{\hangindent=4.35em\noindent\makebox[4.35em][l]{\texttt{U.9.1}}Digital Marketing\par}

{\hangindent=4.35em\noindent\makebox[4.35em][l]{\texttt{U.9.2}}Lead Generation \& Telemarketing\par}

{\hangindent=4.35em\noindent\makebox[4.35em][l]{\texttt{U.9.3}}Marketing, PR \& Brand Strategy\par}

{\hangindent=4.35em\noindent\makebox[4.35em][l]{\texttt{U.10.1}}Language Tutoring \& Interpretation\par}

{\hangindent=4.35em\noindent\makebox[4.35em][l]{\texttt{U.10.2}}Translation \& Localization Services\par}

{\hangindent=4.35em\noindent\makebox[4.35em][l]{\texttt{U.11.1}}AI Apps \& Integration\par}

{\hangindent=4.35em\noindent\makebox[4.35em][l]{\texttt{U.11.2}}Blockchain, NFT \& Cryptocurrency\par}

{\hangindent=4.35em\noindent\makebox[4.35em][l]{\texttt{U.11.3}}Desktop Application Development\par}

{\hangindent=4.35em\noindent\makebox[4.35em][l]{\texttt{U.11.4}}Ecommerce Development\par}

{\hangindent=4.35em\noindent\makebox[4.35em][l]{\texttt{U.11.5}}Game Design \& Development\par}

{\hangindent=4.35em\noindent\makebox[4.35em][l]{\texttt{U.11.6}}Mobile Development\par}

{\hangindent=4.35em\noindent\makebox[4.35em][l]{\texttt{U.11.7}}Other - Software Development\par}

{\hangindent=4.35em\noindent\makebox[4.35em][l]{\texttt{U.11.8}}Product Management \& Scrum\par}

{\hangindent=4.35em\noindent\makebox[4.35em][l]{\texttt{U.11.9}}QA Testing\par}

{\hangindent=4.35em\noindent\makebox[4.35em][l]{\texttt{U.11.10}}Scripts \& Utilities\par}

{\hangindent=4.35em\noindent\makebox[4.35em][l]{\texttt{U.11.11}}Web \& Mobile Design\par}

{\hangindent=4.35em\noindent\makebox[4.35em][l]{\texttt{U.11.12}}Web Development\par}

{\hangindent=4.35em\noindent\makebox[4.35em][l]{\texttt{U.12.1}}Content Writing\par}

{\hangindent=4.35em\noindent\makebox[4.35em][l]{\texttt{U.12.2}}Editing \& Proofreading Services\par}

{\hangindent=4.35em\noindent\makebox[4.35em][l]{\texttt{U.12.3}}Professional \& Business Writing\par}

{\hangindent=4.35em\noindent\makebox[4.35em][l]{\texttt{U.12.4}}Sales \& Marketing Copywriting\par}

\end{multicols}\endgroup

\methodhead{Classifier instructions}

Classify the primary task or requested output across the supplied user messages. Use earlier messages to
interpret short or ambiguous follow-up requests. If multiple tasks appear, choose the primary requested
outcome.

Choose exactly one category from the task list when its content supports the type of activity in the
task. Return its task\_id. Return task\_id U.0.0 (none\_of\_these) when the task is outside the listed
categories or too vague to classify (such as a user simply greeting the model). Never force an unrelated
category merely because the message is work-related.

\Needspace{4\baselineskip}
Independently classify use\_context:

work\_tasks: work for an employer, client, business, professional practice, or income-generating
activity; a clearly professional deliverable -{}- such as a slide deck or email to a client -{}- may imply
work when personal/school cues are absent.

personal\_tasks: personal advice, looking up misc. information, questions regarding household activities,
hobby, relationship advice, or other non-school personal activity. These are messages where the user is
using the AI tool for purposes that seem mostly related to tasks and topics outside of for the purposes
of work.

schoolwork: formal coursework, homework, studying, examination preparation, or a student assignment.
These are messages that seem to be by a student or learner engaging in schoolwork related activities.

unclear: insufficient evidence to distinguish the message as belonging to these purposes.

A recognizable task category and use\_context are separate. For example a student writing code can be
engaged in Web Development plus schoolwork; a household tax question (i.e., advice on filing personal
household taxes) can have the task\_label of Other - Accounting \& Consulting plus use\_context being
personal\_tasks. For messages that have use\_context being work\_tasks, none\_of\_these means other type
of tasks that isn\textquotesingle{}t well captured in the existing list of tasks.

\methodhead{The following are details on each of the tasks}

\begin{itemize}[itemsep=2pt]

\item AI-generated writing, images, audio, or video belongs to the corresponding Writing or Design \& Creative category.

\item AI \& Machine Learning is AI model development/training and development.

\item AI Apps \& Integration is integrating an AI capability/chatbot into an application.

\item Data Analysis \& Testing produces analyses, dashboards, visualization, statistics, or experiments;

\item Data Extraction/ETL gathers, transforms, or processes data;

\item Data Mining \& Management builds pipelines, warehouses, or data-engineering systems.

\item Market Research \& Product Reviews produces customer, product, competitor, or market research.

\item Scripts \& Utilities creates focused automation, scrapers, or scripts; Web Development builds web software/sites; Mobile Development builds mobile apps; Ecommerce Development builds or maintains online stores; ERP/CRM Software configures or builds business-system applications; DevOps \& Solution Architecture deploys or architects operational infrastructure.

\item QA Testing tests software functionality; Information Security \& Compliance secures systems or handles IT controls; Customer Service \& Tech Support are messages requesting answers user/customer issues; Network \& System Administration administers networks, hosts, and systems; Database Management \& Administration administers databases.

\item Web \& Mobile Design produces UX/UI, wireframes, interface prototypes, or site/app visual design; Graphic, Editorial \& Presentation Design produces graphics, presentations, infographics, edited images, or marketing collateral; Branding \& Logo Design produces brand visuals or logos; Product Design concerns creating, designing and developing physical/industrial/fashion/jewelry products; 3D Modeling \& CAD concerns questions around creating or editing CAD models/renderings.

\item Content Writing produces blogs, articles, scripts, creative content, or ordinary editorial copy

\item Sales \& Marketing Copywriting produces persuasive ads, or sales copy or marketing related copy.

\item Professional \& Business Writing produces technical, academic, grant, legal, medical, resume, proposal, or substantive business documents

\item Editing \& Proofreading Services revises existing writing rather than creating completely new writing.

\item Digital Marketing relates to questions around SEO, paid/search/social/email campaigns, or marketing automation

\item Marketing, PR \& Brand Strategy are messages related to defining, ideating or otherwise discussing brand/marketing/content strategy or how to handle PR situations

\item Lead Generation \& Telemarketing are for requests or questions around developing prospects or sales outreach. However creating the written copy itself belongs to Sales \& Marketing Copywriting when copy is the requested deliverable.

\item Project Management is about tasks related to project delivery across domains (such as creating content calendars or tracking tasks and status of projects)

\item Product Management \& Scrum is related to creating, editing, review or otherwise working with product roadmaps, product features or product strategy.

\item Accounting \& Bookkeeping are for user questions related to accounting records or accounts

\item Financial Planning provides financial modeling, planning, or fractional-CFO services

\item Other - Accounting \& Consulting is specifically requests for tax preparation advice, reviews or questions.

\item Finance \& Tax Law provides legal advice about taxes, securities, or financial regulation.

\item Corporate \& Contract Law covers contracts, intellectual property, patents, and business law; Finance \& Tax Law covers financial/tax legal advice; International \& Immigration Law covers immigration/international matters; Public Law covers employment, labor, and regulatory legal matters.

\item Translation \& Localization Services translates written material; Language Tutoring \& Interpretation covers language instruction and live interpretation.

\item Performing Arts are questions related to topics in live performance, acting, singing, or voice talent; Audio \& Music Production are questions related to editing, mixing, or composes audio/music; Video \& Animation are requests to produce, edits, animates, or enhances video.

\item Contract Manufacturing covers supply-chain management, sourcing, procurement, and logistics.

\item Management Consulting \& Analysis covers business analysis and strategy (such as competitive analyses). It also includes professional instructional design and course development.

\item Physical Sciences includes topics related to scientific topics (like biology, chemistry, physics), requests for tutoring in STEM subjects and conducting scientific research; make sure to distinguish actual professional tutoring/research from a student\textquotesingle{}s own schoolwork through use\_context.

\item Community Management \& Tagging includes community moderation and image tagging; Customer Service \& Tech Support concerns general customer service, onboarding, success, and technical support.

\item If multiple categories apply, choose the primary requested deliverable. Do not treat topic/industry alone as a task: a blog post discussing healthcare and medicine blog is Content Writing, medical translation is Translation \& Localization Services, and software for HR is software development rather than Recruiting \& Human Resources.

\end{itemize}

When task\_reasoning and use\_context\_reasoning are requested, explain them briefly without quoting sensitive message text.

\end{classifierprompt}

\clearpage

\section{Summary of AI jobs classification prompt}

\methodhead{Binary classification prompt}

\begin{quote}\begin{classifierprompt}

Overview: You are a job classification assistant.

\noindent Job Information:\par\vspace{-3pt}

\noindent - Post Title: "\{post\_title\}"\par\vspace{-3pt}

\noindent - Post Description: "\{post\_desc\}"\par\vspace{-3pt}

\noindent - Required Job Skills: "\{job\_skill\_string\}"\par\vspace{-3pt}

\noindent - Work Category: "\{work category\}"\par\vspace{-3pt}

\noindent - Work Sub-Category: "\{work sub-category\}"\par\vspace{-3pt}

\noindent - Work Specialization: "\{Work Specialization\}"\par\vspace{-3pt}

\noindent \{conversation\_summary\_if\_available\}\par\vspace{-3pt}

Task: Based on the job title, description, required skills, and the hierarchical categories Work
Category, Work Sub-Category, and Work Specialization, and the summarized conversation context if
available, classify whether this job post is explicitly related to Artificial Intelligence (AI) or not.

Response: Provide a classification of whether the job post is "AI-related" or "Non-AI-related," along with a brief explanation for your reasoning.

\end{classifierprompt}\end{quote}

\methodhead{AI category classification prompt}

\begin{quote}\begin{classifierprompt}

Overview: You are an AI-related job categorization assistant.

\noindent AI-related Job Information:\par\vspace{-3pt}

\noindent - Post Title: "\{post\_title\}"\par\vspace{-3pt}

\noindent - Post Description: "\{post\_desc\}"\par\vspace{-3pt}

\noindent - Required Job Skills: "\{job\_skill\_string\}"\par\vspace{-3pt}

\noindent - Work Category: "\{work category\}"\par\vspace{-3pt}

\noindent - Work Sub-Category: "\{work sub-category\}"\par\vspace{-3pt}

\noindent - Work Specialization: "\{Work Specialization\}"\par\vspace{-3pt}

\noindent \{conversation\_summary\_if\_available\}\par\vspace{-3pt}

Task: For each job, use the job title, description, required skills, and the hierarchical categories Work
Category, Work Sub-Category, and Work Specialization, to classify it into exactly one of the following 8
categories based on best fit with the provided definitions:

\noindent -1. AI Infrastructure and Engineering.\par\vspace{-3pt}

\noindent -2. AI Integration.\par\vspace{-3pt}

\noindent -3. Generative AI \& Creative Production.\par\vspace{-3pt}

\noindent -4. AI Strategy, Consulting \& Productization.\par\vspace{-3pt}

\noindent -5. AI Support, Training \& Labeling.\par\vspace{-3pt}

\noindent -6. AI-Augmented Professional Services.\par\vspace{-3pt}

\noindent -7. Non-AI Work Supporting AI Companies or Projects.\par\vspace{-3pt}

\noindent -8. Other AI Work.\par\vspace{-3pt}

Response: Provide a single AI category, and a brief explanation for your reasoning.

\end{classifierprompt}\end{quote}

\end{document}